\documentstyle[aps,epsfig,floats]{revtex}

\begin{document}
\draft
\title{Critical Behavior of the Three-Dimensional Ising Spin Glass.}

\author{
H.G. Ballesteros$^1$, A. Cruz$^2$, L.A. Fern\'andez$^1$, 
V. Mart\'{\i}n-Mayor$^3$, J. Pech$^2$,\\ J.J. Ruiz-Lorenzo$^4$, 
A. Taranc\'on$^2$, P. T\'ellez$^2$, C.L. Ullod$^2$, 
C. Ungil$^2$.
}
\address{
$^1$ Depto. de F{\'\i}sica Te\'orica, Facultad de CC. F{\'\i}sicas,
Universidad Complutense de Madrid, 28040 Madrid, Spain.\\
%{\tt hector,laf@lattice.fis.ucm.es}\,.\\
$^2$ Depto. de F{\'\i}sica Te\'orica, Facultad de Ciencias,
Universidad de Zaragoza, 50009 Zaragoza, Spain.\\
%{\tt cruz,pech,tarancon,tellez,ungil,clu@sol.unizar.es}\, .\\
$^3$ Dip. di Fisica,
Universit\`a di Roma ``La Sapienza'' and INFN, 
Ple. Aldo Moro 2, 00185 Roma, Italy.\\
%{\tt Victor.Martin@roma1.infn.it}\, .\\
$^4$Depto. de F\'{\i}sica, Facultad de Ciencias, 
Universidad de Extremadura, 06071 Badajoz, Spain.\\ 
%{\tt ruiz@unex.es}\, .
}
\date{\today}

\maketitle


\begin{abstract}
We have simulated, using parallel tempering, the three dimensional
Ising spin glass model with binary couplings in a helicoidal
geometry. The largest lattice ($L=20$) has been studied using a
dedicated computer (the SUE machine).  We have obtained, measuring the
correlation length in the critical region, a strong evidence for a
second-order finite temperature phase transition ruling out other
possible scenarios like a Kosterlitz-Thouless phase transition.
Precise values for the $\nu$ and $\eta$ critical exponents are also
presented.
\end{abstract}

\thispagestyle{empty}

%\newpage

\section{Introduction}

The study of spin-glasses~\cite{BOOKS}, beyond their own physical
relevance, has opened new ways to Statistical Physics. The
solution~\cite{PARISI} of the Sherrington-Kirkpatrick model, that
describes spin-glasses living in infinite dimensions, allowed the
introduction of a new set of ideas that have found applications in
very different contexts, like Optimization, Neural Networks, and so
on. Yet the applicability of the rich infinite-dimensional physical
picture to describe the low-temperature physics of three-dimensional
spin-glass materials (like, for instance, ${\mathrm Cu\,Mn}$,
${\mathrm Ag\,Mn}$, ${\mathrm Eu}_x\,{\mathrm Sr}_{1-x}\,{\mathrm S}$,
see~\cite{EXPBOOK}) is still controversial~\cite{CONTROVERSIAL}.
Furthermore, a rather simpler question, {\em what is the nature of the
spin-glass phase transition?}, has not yet found a fully satisfactory
answer. Although the very existence of a phase transition has been
questioned, from the experimental side, there is now a wide consensus
on its existence, as signalled by the behavior of the non-linear
susceptibility~\cite{EXP}.

On the other hand, the theoretical approach is almost limited to the
Monte Carlo simulation of the Edwards-Anderson model, given the
enormous difficulties found when using field-theoretic renormalization
group techniques~\cite{deDOMINICIS}. Recent numerical
simulations~\cite{KAYOUNG,INPARU,MAPARU,INMAPARU,BERGJANKE} have found
indications of a finite-temperature phase transition, which has been
confirmed in Ref.~\cite{PACA}. However, the
possibility~\cite{INMAPARU} of a Kosterlitz-Thouless like phase
transition (an exponential divergence of the correlation-length at the
critical temperature followed by a line of critical points) could {\em
not} be excluded~\cite{PACA}. Even so, critical-exponent estimates
that could be compared with experiments were
obtained~\cite{KAYOUNG,INPARU,MAPARU,INMAPARU,BERGJANKE,PACA,OGIELSKI,BY1,BY2}
by assuming power-law divergences at the critical temperature (i.e.,
{\em non} Kosterlitz-Thouless behavior).  However, the statistical errors of
those estimates (10\% for the correlation-length exponent $\nu$, and
$15$ \% in the anomalous dimension, $\eta$), and that of the critical
temperature estimate seems poor compared to similar computations for
ordered systems, which is due to the numerical difficulties
encountered on the simulation of the Edwards-Anderson model. In fact,
the issue (crucial for accurate calculations of critical exponents) of
the scaling corrections has not been addressed in previous works,
exception made of Ref.~\cite{PACA}.

In this work, we shall perform a detailed study of the critical
behavior of the Edwards-Anderson model. The numerical simulations have
been in part performed on a dedicated computer (the SUE machine, see
below for more details), on which we have been able to thermalize 6920
samples of $20^3$ lattices at the critical temperature, the largest
thermalized lattices in previous studies at similar temperatures being
$16^3$.  For the thermalization deep inside the critical region we
have used the Monte Carlo exchange method (also known as parallel
tempering)~\cite{PARTEMP,ENZO,BOOK}.

Our study shares with Ref.~\cite{PACA} the definition of the
finite-lattice correlation-length~\cite{COOPER}, and a heavy use of
the Finite-Size Scaling (FSS) Ansatz~\cite{FSSBOOK}. Yet, both
analyses are rather different. Ref.~\cite{PACA} uses the techniques of
Ref.~\cite{EXTRA} to extrapolate the measures taken on lattices which
are small compared with the correlation-length, to the thermodynamic
limit. On the other hand, we use the quotient method~\cite{QUOTIENT},
where measures taken on two lattices are compared at the temperature
at which the correlation-length measured in units of the lattice-size
coincides for both.

For the particular problem of spin-glasses, the method of
Ref.~\cite{PACA} has the advantage of not requiring the thermalization
of large lattices at large correlation-lengths (the dynamical critical
exponent for the three dimensional Ising spin glass in the critical
region is near $7$~\cite{ZETA}).  On the other hand, the quotient
method offers the possibility of extremely precise determinations of
critical exponents and temperatures, and a rather transparent control
of scaling-corrections, also in disordered systems~\cite{DISORDERED}.
The main drawback for its use on spin-glass systems is that it
requires measures taken on several pairs of lattice of widely
different sizes at the critical temperature, which is rather difficult
due to the above mentioned thermalization problems\cite{FCC}.

We obtain very precise estimates for critical exponents which are
compared with the estimates of other groups and with available
experimental results. The issue of scaling-corrections will be
discussed, and a rough estimate of $\omega$ (the correction-to-the-scaling
exponent) will be obtained.

An additional bonus of our computational strategy is that high-quality
data for the spin-glass correlation length are generated on large
lattices at the critical region. This allows for a detailed comparison
with the archetypical model displaying a Kosterlitz-Thouless phase
transition, the XY model in two dimensions (for which a cluster method
is available, making the simulation almost costless). The finite-size
scaling behavior of both models is not only quantitatively but
qualitatively different. We therefore present strong evidence against
Kosterlitz-Thouless behavior on the Edwards-Anderson model.

Finally, we also consider the question of the appropriate cumulant for
the study of the spin-glass phase transition. In
Ref.~\cite{FELIXMARTA} it has been argued that the Binder
cumulant~\cite{BINDER} works poorly, in marked contrast with ordered
systems. It is also claimed that the cumulant $G$ introduced in
Ref.~\cite{MANAPAPIRIZU} for the study of systems without
time-reversal symmetry, see Eq.(\ref{GFETEN}), does a better job. We
shall show that it suffers from similar scaling corrections but of
opposite {\em sign}, so that its crossings happens at temperatures
higher than the critical point. This is rather advantageous from the
point of view of thermalization. On the other hand, its measures are
far noisier than the ones of the finite-lattice correlation-length,
and it also suffers from stronger corrections to scaling.

The large statistics needed to obtain precise results on larger
lattices has been possible by the use of a dedicated computer based on
programmable components. Details about the machine can be seen
in Refs.~\cite{CLU,RUCLU,SUE}. SUE, for Spin Update Engine,
consists of twelve boards attached to a PC.  Each of these boards
contains programmable devices and memories, allowing the simulation of
eight lattices of size up to $L=60$.  The presence in each board of a
component dedicated to random number generation allows the use of the
Heat Bath algorithm. During the simulation, Schwinger-Dyson
equations~\cite{SD,SDE} were used to check for the correctness of the
random-number sequences.  Periodically, the spin configurations are
downloaded to the PC, where the measures are performed and stored.
When using the parallel tempering scheme, the PC controls the
mechanism, interchanging the configurations corresponding to adjacent
temperatures when appropriate.  The update speed of the whole system
is $0.22$ ns/spin, one hundred times faster than one Alpha EV5, 600
MHz processor running multi-spin code.

The layout of the rest of this paper is as follows. In the next
section we introduce the model, the definition of the observables and
we review the basis of our Finite Size Scaling (FSS) method. The
third section is devoted to analyze the statistical quality of our
data, in particular we discuss the thermalization, the parallel
tempering parameters and our choice of the number of samples versus
number of steps inside one sample. The discussion of our numerical 
results follows, where we compare them with previous numerical
simulations and experiments. We end the paper with the conclusions.

\section{Model, observables  and the FSS method}

We have studied the three dimensional Ising spin glass defined on a
cubic lattice ($L\times L \times L$) with helicoidal boundary
conditions~\cite{HELICOIDAL}, whose Hamiltonian is
\begin{equation}
{\cal H}=-\sum_{<i,j>} \sigma_i J_{ij} \sigma_{j} \ .
\end{equation}
The volume of the system is $V=L^3$, $\sigma_i$ are Ising variables,
$J_{ij}$ (uncorrelated quenched disorder) are $\pm 1$ with equal
probability, and the sum is extended to all pairs of nearest
neighbors. 
Due to the quenched nature of the disorder, one needs to  perform first
the thermal average for a given configuration of the
$J_{ij}$ (denoted by $\langle\cdots\rangle$ hereafter),
and later the average over the disorder realization (that will be
indicated by an overline).  The choice of helicoidal boundary
conditions is mandatory (for us) because the hardware of the SUE
machine has been optimized for them.

We have simulated the smaller lattice sizes ($L=5$ and $10$) 
in  parallel machines built of Pentium-Pro processors (the
RTNN machines) using multi-spin coding.  
We have checked that the $L=10$ and $L=5$ lattices are properly
thermalized with a standard Heat Bath method (without parallel tempering).
The larger lattice ($L=20$) has been simulated in the SUE
machine using parallel tempering and Heat Bath.

We shall describe in depth the thermalization test and the total
statistics achieved in the next section.

\subsection{Observables}

It is well known that observables in spin-glasses need to be defined
in terms of real replicas, that is, for every disorder realization,
one considers two thermally independent copies of the system
$\{s^{(1)}_i,s^{(2)}_i\}$.  Observables are most easily defined in
terms of a spin-like field, the so-called overlap field:
\begin{equation}
q_i= s^{(1)}_is^{(2)}_i\,.
\end{equation}
The total overlap is the lattice average of the $q_i$
\begin{equation}
q=\frac{1}{V} \sum_i q_i\,,
\end{equation}
while the (non-connected) spin-glass susceptibility is
\begin{equation}
\chi_q= V \overline{\langle q^2 \rangle }\, .
\label{chi}
\end{equation}
In Finite-Size Scaling studies, it is useful to have dimensionless
quantities, that go to a constant
value at the critical temperature.  A standard example of such a
quantity is the Binder cumulant
\begin{equation}
g_4=\frac{3}{2}- 
\frac{1}{2}
\frac{\overline{\langle q^4  \rangle }} 
{\overline{\langle q^2  \rangle}^2 } \, .
\label{binder}
\end{equation}
Other example is the $g_2$ cumulant~\cite{DISORDERED}, that measures
the lack of self-averageness of the spin-glass susceptibility
\begin{equation}
g_2=\frac{\overline{\langle q^2\rangle^2}-\overline{\langle q^2\rangle}^2}
{\overline{\langle q^2\rangle}^2}\,.
\end{equation}
Of course, any smooth function of these two dimensionless quantities,
$g_2$ and $g_4$, is dimensionless itself. In Ref.~\cite{MANAPAPIRIZU}
it was proposed to study the cumulant $G$
\begin{equation}
G=\frac{g_2}{2-2g_4}\,
\label{GFETEN}
\end{equation}
because it exhibits a significant reduction of scaling corrections.
It has also been argued in Ref.\cite{FELIXMARTA} that $G$ can be
extremely helpful for the characterization of the spin-glass phase,
but this issue is out of the scope of this work.  However all the
above defined dimensionless quantities, $g_4$, $g_2$ and $G$, require
the evaluation of a four-point correlation function, which is
statistically a much noisier quantity than a two-point one.  One
observable of this kind is the correlation-length, which is defined in
terms of the two-point correlation function, and its quotient with the
lattice size is again dimensionless. We therefore are faced with the
problem of defining a correlation-length on a finite lattice. In
Ref.~\cite{COOPER} it was shown how to do it, from the following
considerations. Let us call $C(\mbox{\boldmath$r$})$ the correlation
function of the overlap field,
\begin{equation}
C({\mbox{\boldmath$r$}})=\frac{1}{V}\sum_i \overline{\langle q_i
q_{i+{\mbox{\scriptsize \boldmath$r$}}}\rangle}
\end{equation} 
and $\hat C(\mbox{\boldmath$k$})$ its Fourier transform.  Notice that the
spin glass susceptibility is simply $\hat C(0)$.  Then, inside the
critical region on the paramagnetic side and in the thermodynamical
limit, one has
\begin{eqnarray}
\hat C({\mbox{\boldmath$k$}})&\propto&\frac{1}{k^2+\xi^{-2}}\ ,\quad 
\Vert{\mbox{\boldmath$k$}}\Vert\ll \xi^{-1}\,,\\
\xi^{-2}&=&-\frac{1}{\hat C}\left.
\frac{\partial\hat C}{\partial {\mbox{\boldmath$k$}}^2}
\right|_{{\mbox{\scriptsize\boldmath$k$}}^2=0}\label{XIDERIVADA}\,.
\end{eqnarray}

On a finite lattice, the momentum is discretized, and one
uses~\cite{COOPER} a finite-differences approximation to
Eq.(\ref{XIDERIVADA}),
\begin{equation}
\xi^2=\frac{1}{ 4  \left[ \sin^2(k^x_{\mathrm m}/2)+\sin^2(k^y_{\mathrm m}/2)
+\sin^2(k^z_{\mathrm m}/2)\right]} 
\left[ \frac{\chi_q}{\hat C({\mbox{\boldmath$k$}}_{\mathrm m})}-1 \right] \ ,
\label{correlation}
\end{equation}
where $\chi_q$ was defined in Eq.~(\ref{chi}) and 
$\mbox{\boldmath$k$}_{\mathrm m}$ is the minimum wave-vector allowed 
for the used
boundary conditions, which in our case is ${\mbox{\boldmath$k$}}_{\mathrm
m}=(2\pi/L, 2 \pi /L^2, 2\pi/L^3)$. Of course, Eq.(\ref{XIDERIVADA})
holds on the thermodynamic limit ($L\gg \xi$) of the paramagnetic
phase. As we do not use connected correlation functions, $\xi$ has
sense as a correlation length only for $T>T_{\mathrm c}$.

We can study the scaling behavior of the finite-lattice
definition~(\ref{correlation}) on a critical point, where the
correlation function decays (in $D$ dimensions) as
$r^{-(D-2+\eta)}$. The behavior of the Fourier transform of the
correlation function for large $L$ in three dimensions is given by
\begin{equation}
\hat C(k)\sim \int_0^L d r\, r^{1-\eta}\frac{\sin(k r)}{kr}\,,
\end{equation}
and one finds that $\chi_q/\hat C({\mbox{\boldmath$k$}}_{\mathrm m})$ goes to
a constant value, larger than unity, because 
$\Vert{\mbox{\boldmath$k$}}_{\mathrm m}\Vert={\cal O}(1/L)$.  
Furthermore, $\xi/L$ tends to a
universal constant at a critical point (like the Binder cumulant
$g_4$).  Moreover, on a broken-symmetry phase, where the fluctuations
of the order parameter are not critical, one has $\chi_q={\cal O}(V)$,
while $\hat C({\mbox{\boldmath$k$}}_{\mathrm m})={\cal O}(1)$. Therefore the
full description of the scaling behavior of $\xi/L$ is as follows.
Let $\xi_\infty$ be the correlation-length on the infinite lattice: in
the paramagnetic phase, for $L\gg \xi_\infty$, one has $\xi/L={\cal
O}(1/L)$. On the FSS region, where $\xi_\infty\geq L$, $\xi/L={\cal
O}(1)$, while on a broken-symmetry phase on a lattice larger than the
scale of the fluctuations, $\xi/L={\cal O}(L^{D/2})$. Consequently, 
if one plots $\xi/L$ for several lattice sizes
as a function of temperature, the different graphics will cross at the
critical temperature.

Finally, a very useful quantity is the value of the Hamiltonian,
which is used to measure the derivatives of a generic observable, $O$, 
with respect to the inverse temperature $\beta$ 
\begin{equation}
\partial_\beta {\langle O\rangle}=
{\langle O {\cal H}^{(1)}+ O{\cal H}^{(2)}\rangle}-
{\langle O\rangle\langle {\cal H}^{(1)}+{\cal H}^{(2)} \rangle}\,,
\label{DERICONEXA}
\end{equation}
where the $^{(1)}$ and $^{(2)}$ superscripts refers to the two
replicas needed to construct the operator $O$. Similarly one
generalizes the standard reweighting method~\cite{FALCIONI}, that
allows to extrapolate the measures taken at $\beta$ to neighboring
values:
\begin{equation}
\langle O\rangle_{\beta+\Delta\beta}=
\frac
{\langle O\, 
{\mathrm e}^{\Delta\beta{\cal H}^{(1)}+\Delta\beta{\cal
H}^{(2)}}\rangle_\beta}
{\langle 
{\mathrm e}^{\Delta\beta{\cal H}^{(1)}+\Delta\beta{\cal H}^{(2)} }\rangle_\beta}
\,.\label{REWEIGHTING}
\end{equation}

\subsection{The FSS method}

We have used the quotient method~\cite{QUOTIENT}, in order to compute
the critical exponents. We recall briefly the basis of this
method. Let $O$ be a quantity diverging in the thermodynamical limit
as $t^{-x_O}$ ($t=T/T_{\mathrm c}-1$ being the reduced
temperature). We can write the dependence of $O$ on $L$ and $t$ in the
following way~\cite{FSSBOOK}
\begin{equation}
O(L,t)=L^{x_O/\nu} \left[F_O\left(\frac{L}{\xi(\infty,t)}  \right)+
{\cal O}(L^{-\omega},\xi^{-\omega}) \right] ,
\label{obs}
\end{equation}
where $F_O$ is a (smooth) scaling function and $(-\omega)$ is the
corrections-to-scaling exponent (e.g., $\omega$ is the largest
negative eigenvalue of the Renormalization Group transformation). This
expression contains the not directly measurable term $\xi(\infty,t)$,
but if we have a good definition of the correlation length in a finite
box $\xi(L,t)$, Eq.~(\ref{obs}) can be transformed into
\begin{equation}
O(L,t)=L^{x_O/\nu} \left[G_O\left(\frac{\xi(L,t)}{L}  \right)+
{\cal O}(L^{-\omega}) \right] ,
\end{equation}
where $G_O$ is a smooth function related with $F_O$ and $F_\xi$ and
the term $\xi_\infty^{-\omega}$ has been neglected because we are
simulating deep in the scaling region.  We consider the quotient of
measures taken in lattices $L$ and $sL$ at the same temperature
\begin{equation}
Q_O(s,L,t)=\frac{O(sL,t)}{O(L,t)}\,.
\end{equation}
Then, 
the main formula of the quotient method is 
\begin{equation}
\left.Q_O\right|_{Q_\xi=s}=
s^{x_O/\nu}+{\cal O}(L^{-\omega})\ ,
\label{QUO}
\end{equation}
i.e., we compute the reduced temperature $t$, at which the correlation
length verifies $\xi(sL,t)/\xi(L,t)=s$ and then the quotient between
$O(sL,t)$ and $O(L,t)$. In particular, we apply formula
(\ref{QUO}) to the overlap susceptibility, $\chi_q$, and the
$\beta$-derivative of the correlation length, $\partial_\beta\xi$,
whose associated exponents are:
\begin{eqnarray}
x_{\partial_\beta\xi}&=&1+\nu\,,\\ 
x_{\chi_q}&=&(2-\eta)\nu .
\label{OBSERVABLES}
\end{eqnarray}
Notice that $\left.Q_O\right|_{Q_\xi=s}$ can be measured with great
precision because of the large statistical correlation between $Q_O$
and $Q_\xi$. It is very important that in order to use Eq.(\ref{QUO})
one does not need the infinite-volume extrapolation for the critical
temperature but, instead, a reweighting method~\cite{FALCIONI} is
crucial to fine-tune the $Q_\xi=s$ condition.  From
Eq.(\ref{QUO}) one directly extracts {\em effective} exponents
(i.e., lattice-size dependent) and later on checks for
scaling-corrections. Another advantage of the quotient method is that
the crossing temperature for $\xi/L$ (i.e., the temperature for which
$Q_\xi=s$) scales as
\begin{equation}
t^{{\mathrm crossing}}_{s,L}\propto L^{-\omega-1/\nu}\,,
\end{equation} 
so that one works a factor $L^{-\omega}$ closer to the critical point
than with other FSS methods, such as measuring at the maximum of (say)
the connected susceptibility.  This puts considerably less stress on
the quality of the reweighting method. It should also be mentioned
that one could modify Eq.(\ref{QUO}), and measure at the crossing
point of a cumulant such as $g_4$, $g_2$ or $G$. The results should
coincide, up to scaling corrections, but given the better statistical
quality of the measures of $\xi/L$, the error bars would significantly
grow.  

\section{Statistical Quality of the data}

When designing a simulation for a disordered model, one needs to
carefully consider how many measures will be taken on each sample,
$N_{\mathrm I}$, and the number of samples to be simulated
$N_{\mathrm S}$.  Two competing effects need to be balanced for
this. On the first place, from the error analysis of a generic
observable, $O$, one has ($\eta$ is a Gaussian number of zero-mean and
unit variance)
\begin{equation}
\left(\overline{\langle O \rangle}^{{\mathrm Monte Carlo}}\ -\ 
\overline{\langle O \rangle}\right)^2=\frac{\eta}{N_{\mathrm S}}
\left(\sigma_{{\mathrm S},O}^2+
\frac{2\tau_O\sigma_{{\mathrm I},O}^2}{N_{\mathrm I}}\right)\,,
\label{ERRORANALISIS}
\end{equation}
where $\sigma_{{\mathrm S},O}$ is the variance between different
samples of the exact thermal averages, $\sigma_{{\mathrm I},O}$ is the
disorder-averaged variance for the measures on a sample, and $\tau_O$
is an averaged (integrated) autocorrelation time~\cite{SOKAL}.  This
shows that the optimum value of $N_{\mathrm I}$ cannot be much greater
than $2\tau_O\sigma_{{\mathrm I},O}^2/\sigma_{\mathrm S}^2$.  On the
other hand, when evaluating non-linear functions of thermal-averages,
as in Eqs.(\ref{DERICONEXA},\ref{REWEIGHTING}), a bias of order
$2\tau_O/N_{\mathrm I}$ is present (for the reweighted measures a bias
polynomial in $2\tau_O/N_{\mathrm I}$ is expected). If
$\sqrt{N_{\mathrm S}}$ is not much smaller than $N_{\mathrm I}$, the
statistical errors will shrink enough as to uncover the bias, and we
have two conflicting goals for the optimization of $N_{\mathrm I}$ and
$N_{\mathrm S}$. In order to solve the dilemma, we have followed the
same procedure as in Ref.~\cite{DISORDERED} to eliminate the bias. One
first evaluate the non-linear function with the full set of data, then
divides the data in two sets of length $N_{\mathrm I}/2$ for each of
which the function is evaluated and the two results are averaged, and
the procedure is repeated dividing the data in four sets of
$N_{\mathrm I}/4$ measures. We thus have three estimates of the
non-linear function, with bias of order $1/N_{\mathrm I}$,
$2/N_{\mathrm I}$ and $4/N_{\mathrm I}$ respectively. The three
estimates are used in a quadratic (in $1/N_{\mathrm I}$) extrapolation
to $1/N_{\mathrm I}=0$, which is later on averaged over the disorder.
In order to have meaningful results from this extrapolation, it is
crucial that $2\tau/N_{\mathrm I}$ be a reasonably small number.  The
values of $N_{\mathrm I}$ and $N_{\mathrm S}$ for our simulations are
shown in table~\ref{table:par}. We remark that we need also to balance
the Heat-bath steps, done by the dedicated machine SUE, and the
parallel tempering steps done by the PC which handles SUE (during this
time SUE is stopped).

\begin{table}[t!]
\centering
\begin{tabular}{rrrr}% \hline
\multicolumn{1}{c}{$L$}&\multicolumn{1}{c}{$N_{\mathrm S}$}
&\multicolumn{1}{c}{$N_{\mathrm HB}$}&\multicolumn{1}{c}{$N_{\mathrm I}$}\\ 
\hline\hline
5   & 40 000   &  200 000          & 200        \\ \hline
10  & 40 000   &  500 000          & 500     \\\hline
20  & 6 920    &  6 553 600         & 400   % \\\hline
\end{tabular}
\caption{
\protect\label{table:par} Statistics used. For each sample, we run 
$N_{\mathrm HB}$ Heat Bath sweeps and perform $N_{\mathrm I}$ measures.
Each sample is previously thermalized with $N_{\mathrm HB}$ iterations.
In the $L=20$ lattice, we carry out a parallel tempering step after each
measure.}
\end{table}

With our simulation strategy ($N_{\mathrm I}\ll N_{\mathrm S}$), it is
crucial to check that the system is sufficiently thermalized while
taking measures. A very efficient algorithm for thermalizing
spin-glass systems is parallel tempering~\cite{PARTEMP,ENZO,BOOK}.  In
order to obtain an efficient parallel tempering we must select a range
of $\beta$ values, and the number of intervals in this range.  The
range is fixed in the following way: The faster decorrelation time is
at the lowest $\beta$; as we run a fixed number of iteration between
parallel tempering sweeps, this number must be greater than the
autocorrelation time at this $\beta$ value. For these values of
$\beta$, away from the transition point, the correlation length $\xi$
is (almost) $L$-independent. Running around $10^4$ sweeps and
considering that the correlation time grows as $\xi^7$~\cite{ZETA} we
use finally $\beta_{\mathrm min}= 0.70$. The largest value must be a
bit over the crossing point, which we had estimated previously around
0.88. We use then $\beta_{\mathrm max}= 0.92$.  The number of $\beta$
values is fixed by controlling that the probability of changes in the
parallel tempering is significant.  This number depends on $L$ and for
$L=20$ we have used $12$ values of $\beta$, obtaining a probability of
transition around $30\%$.  All the systems are a significant time in
the lower $\beta$ values, where decorrelation is faster.

\begin{figure}[ht!]
\begin{center}
\leavevmode
\epsfig{file=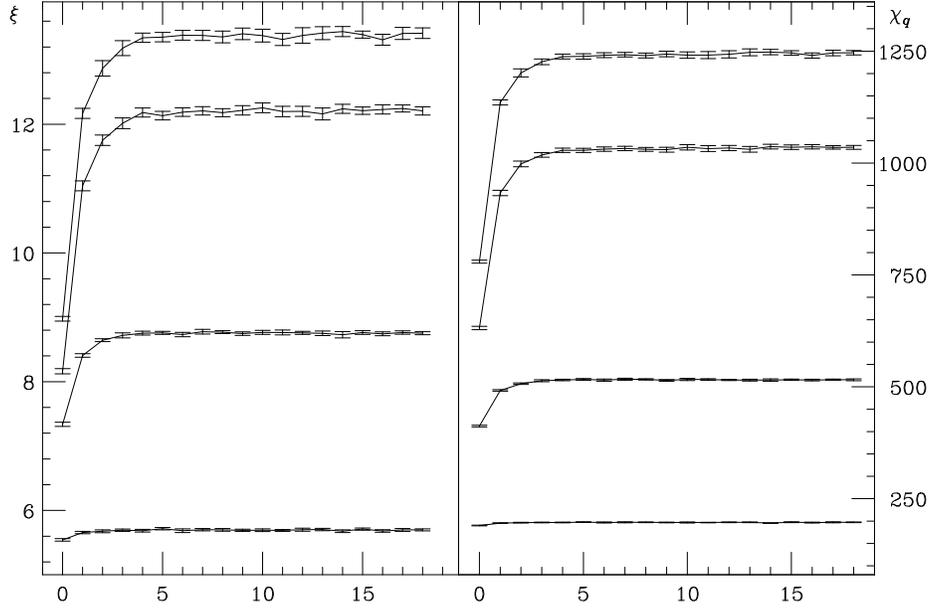,width=0.5\linewidth,angle=90}
\end{center}
\caption{Mean values of the susceptibility and correlation length in the
$L=20$ lattice for successive bins of 40 measures, for several $\beta_i$
values in $L=20$ (from top to bottom they correspond to $\beta_{11}$,
$\beta_9$, $\beta_4$ and $\beta_0$, respectively). We remark that
$\beta_{11}=\beta_{\mathrm max}=0.92$ corresponds with our coldest temperature.
In the following, we discard the first 10 bins and
average the remaining 10.}
\label{TERMA}
\end{figure}
A first thermalization check is summarized in Fig.~\ref{TERMA}.  The
measures taken on a sample are divided in twenty blocks, and the
correlation-length and the spin-glass susceptibility are calculated
with these blocks. No thermalization bias can be resolved after the
fifth block at the lowest temperature. However, the first ten blocks
have been discarded for safety.

\begin{figure}[t!]
\begin{center}
\leavevmode
\epsfig{file=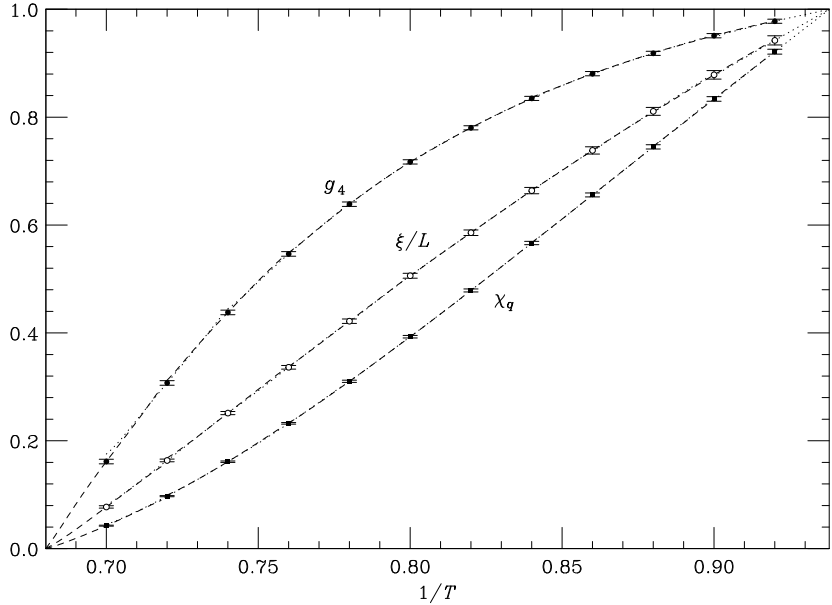,width=0.5\linewidth,angle=90}
\end{center}
\caption{Correlation length, susceptibility and $g_4$ cumulant
for the $L=20$ lattice, as obtained from
the extrapolation method of Eq.(\ref{REWEIGHTING}), corrected for
bias. We show the results for all the twelve $\beta$ values of the
simulation (point with error bars) together with the extrapolation
(alternatively shown in dashed and dotted lines). The curves have
been linearly scaled to fit in the figure.}
\label{ESTADISTICA}
\end{figure}

Yet the results in Fig.~\ref{TERMA} do not really show that we are
collecting a reasonable number of measures on each sample (so that
$2\tau/N_{\mathrm I}$ is small), because the time needed to obtain a
thermalized measure is not straightforwardly linked to the time needed
to obtain an independent measure. In fact, the latter is related with
the time needed to overcome free-energy barriers, in order to visit
different relevant regions of phase-space~\cite{BERGJANKE}, while the
former is related with the time needed to reach {\em at least one}
relevant region of phase space, because the huge number of samples
avoids a biased estimation. A better test comes from the smoothness of
the bias-corrected reweighting extrapolation. In
Fig.~\ref{ESTADISTICA} we show the extrapolated values $\xi/L$, $g_4$
and $\chi_q$ from each of the $\beta_i$ of the parallel tempering
(alternatively in dotted and dashed lines). The mismatch between
different extrapolations is much smaller than the error bars, which is
due to the fact that the same samples are being simulated for all
$\beta_i$ values. In fact, for $N_{\mathrm I}\to\infty$ the mismatch
would completely disappear, while the statistical error bars would not
be smaller unless $N_{\mathrm S}$ grows (see
Eq.~(\ref{ERRORANALISIS})).

Finally, along the simulation the Schwinger-Dyson equations have been
checked finding a perfect agreement. For instance, the equation in
\cite{SDE} holds within a 0.04\% for the $L=20$ lattice.

\section{Numerical results}

Once a set of measures of the finite-lattice correlation-length is at
our disposal, the first question one can answer regards the nature of
the spin-glass phase transition in three dimensions. Indeed, if the
Kosterlitz-Thouless scenario was realized, $\xi/L$ on the $L\to\infty$
limit, would be zero for temperatures higher than the critical one,
and then it would abruptly jump at the critical temperature to a
finite value. For lower temperatures, since the system would still be
critical, it would keep a finite value, presumably growing with
lowering temperatures. On a finite lattice, $\xi/L$ would be a
decreasing function of $L$ on the paramagnetic phase, and in the
critical region (from the critical temperature to lower ones), it
would have a $L$ independent value, according to the FSS Ansatz, up to
scaling corrections. Therefore, in the most economic scenario, where
scaling corrections are small, the different $\xi/L$ curves do not
cross, but simply merge in the low-temperature region.

\begin{figure}[t!]
\begin{center}
\leavevmode
\epsfig{file=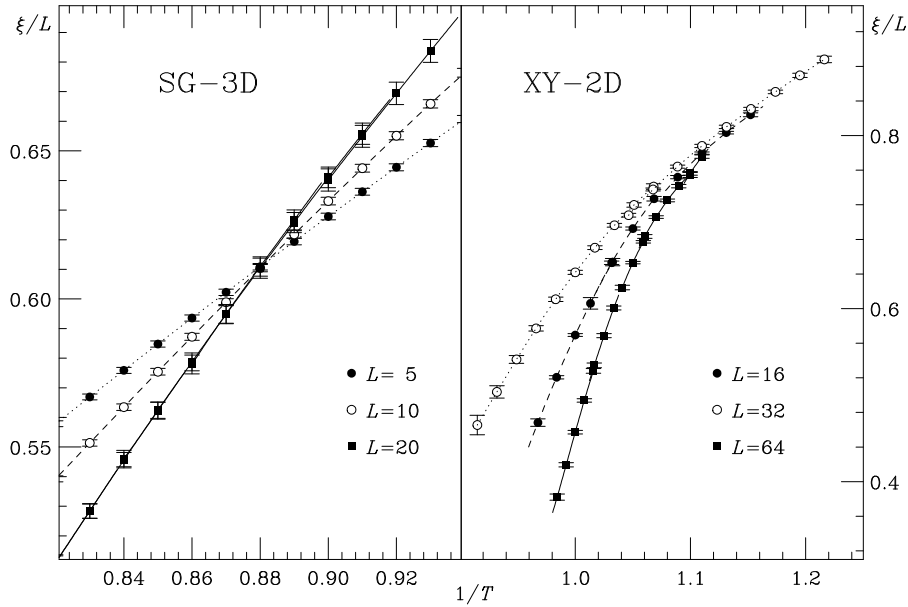,width=0.5\linewidth,angle=90}
\end{center}
\caption{Correlation length in units of the lattice size for
the 3D Edwards-Anderson model (left), 
and for the 2D XY-model that displays a Kosterlitz-Thouless
phase transition.}
\label{DOSXIS}
\end{figure}

In Fig.~\ref{DOSXIS}, we plot $\xi/L$ for the Edwards-Anderson model
in three dimension and for the XY model in two dimensions (whose
simulation is almost costless in computer time~\cite{SIM}).  We see
that while the XY model follows quite closely the above sketched
behavior, the Edwards-Anderson model has a very neat
crossing. Therefore, one may conclude that the Kosterlitz-Thouless
scenario is ruled out by the data, unless scaling-corrections of a
very exotic nature would be present.
Notice, that for the two dimensional XY model, $\xi/L$ and the Binder
cumulant behave in the same way with $L$ and $T$~\cite{INMAPARU}.

\begin{figure}[ht!]
\begin{center}
\leavevmode
\epsfig{file=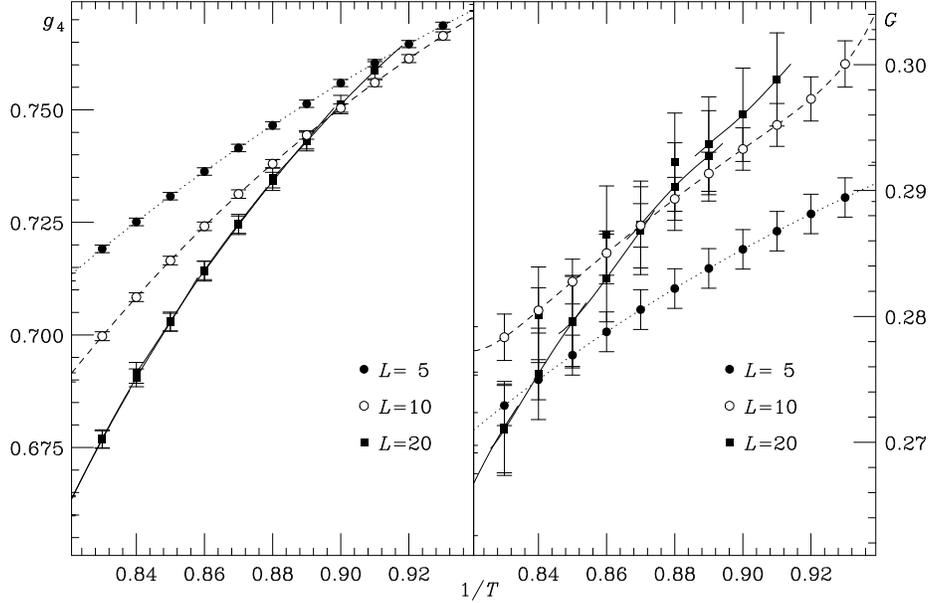,width=0.5\linewidth,angle=90}
\end{center}
\caption{Cumulants $g_4$ and $G$ for lattices $L=5,10,20$ versus 
inverse temperature.}
\label{FIGGFETEN}
\end{figure}

This is an interesting point to compare the behavior of the $g_4$ and $G$
cumulants, defined above, with $\xi/L$. 
From Fig.~\ref{FIGGFETEN}, it is 
clear that the measures for $G$ are much noisier than for
$\xi/L$. Moreover, also the scaling corrections are larger, as
made evident by the large shift between the crossing of the $5$ and
$20$ lattice, and the crossing of the $10$ and $20$ lattice. The
scaling corrections for $G$ and $g_4$ are of opposite signs, so
that one can safely conclude that the real critical point is bracketed
by this two sets of crossing-points.

% exponents
\begin{table}[b!]
\centering
\begin{tabular}{cddd}% \hline
$L$ & $T_{\mathrm c} $& $\nu$ & $\eta$\\ \hline\hline
(5,10)  & 1.134(9) & 2.39(7)        &--0.353(9)    \\\hline
(10,20)  & 1.138(10)& 2.15(15)       &--0.337(15)   %\\\hline
\end{tabular}
\caption[0]{
\protect\label{table:nu} 
Critical exponents computed from the crossing points of $\xi/L$ for
$(L,2L)$ pairs.}
\end{table}

For the critical exponents, our results found using Eq.~(\ref{QUO})
are displayed in table~\ref{table:nu}. Finite-size scaling corrections
cannot be resolved within errors, specially if one realizes that the
results for the $(5,10)$ pair are anticorrelated with the results for
the $(10,20)$ pair (the measure in $L=10$ appear once in the numerator
and the other time in the denominator of Eq.(\ref{QUO})).  We take as
our final estimate our results for the $(10,20)$ pair, that can be
compared with the most recent experimental measures and with other
numerical calculations displayed in table~\ref{table:sim}. It would be
very interesting to check that systematic errors due to finite-size
effects are smaller than the statistical ones. This would require to
parametrize the scaling corrections, and therefore to have precise
measures on a wide range of lattice sizes. Unfortunately, the huge
dynamical critical exponent of the available simulation algorithms for
this model makes extremely difficult to thermalize, at the critical
point, lattices larger than $L=20$ with present technological
capabilities.

Nevertheless, it is possible to obtain information about $\omega$
studying the quotients of cumulants $Q_g$ at the point where $\xi/L$
crosses. At this point, $Q_g$ should be 1 up to scaling corrections 
and can be parameterized as
\begin{equation}
Q_g=1+A_g L^{-\omega}\,.\label{LINEAR}
\end{equation}
Before presenting our results for $\omega$, let us recall the values
obtained by Palassini and Caracciolo~\cite{PACA} working in the
thermodynamical limit. They computed two different scaling-corrections
exponents, $\Delta$ and $\theta$. These exponents are obtained form
the asymptotic formulae that hold in the scaling region in the
thermodynamic limit, in the paramagnetic side
\begin{equation}
\chi=A~\xi^{2-\eta}(1+{\cal O}(\xi^{-\Delta}))\, , \,\,\,
\xi=B ~|t|^{-\nu}(1+{\cal O}(|t|^{\theta}))\,.
\end{equation}
One can check from the above expressions, that
$\Delta=\omega$ and $\theta=\omega \nu$. 
From their values of $\Delta$, $\theta$ and $\nu$ one readily obtains
\begin{equation}
\omega(\Delta)=1.3\scriptsize\begin{array}{c}
+0.2\\-0.3\end{array}\,\, ,\,\,\, \omega(\theta,\nu)=0.78
\scriptsize\begin{array}{c}
+0.47\\-0.28\end{array}\,.\label{OMEGAPACA}
\end{equation}

\begin{figure}[t!]
\begin{center}
\leavevmode
\epsfig{file=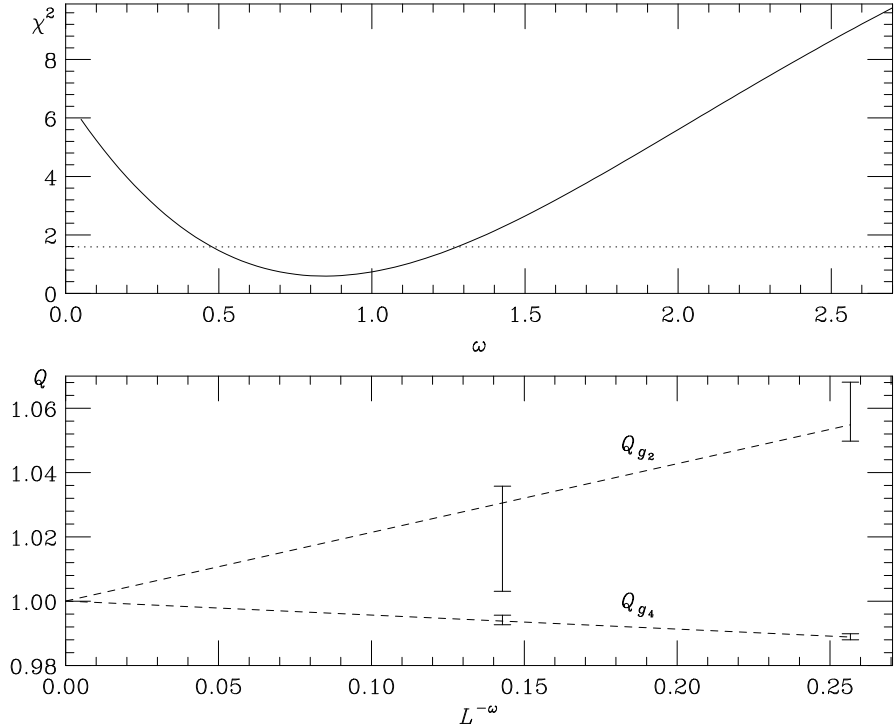,width=0.6\linewidth,angle=90}
\end{center}
\caption{Quotients of the cumulants $g_2$ and $g_4$ for pairs $(L,2L)$
as functions of $L^{-\omega}$ (lower figure). 
The corrections-to-scaling exponent
has been obtained minimizing the $\chi^2$ function (see (\ref{LINEAR}))
using the full covariance matrix. The horizontal dotted line (in the upper
figure) is given by the value of $\chi^2$ at the minimum plus one.}
\label{OMEGA}
\end{figure}

We have fitted our measures for $g_4$ and $g_2$ following
(\ref{LINEAR}). We have 4 points for adjusting 3 parameters, but we
obtain the value
\begin{equation}
\omega=0.84\scriptsize\begin{array}{c} +0.43\\-0.37\end{array}\,,
\label{NUESTROOMEGA}
\end{equation}
with $\chi^2=0.6$. See Fig.~\ref{OMEGA} for more details. Although the
statistical error is rather large, the compatibility of our result
with those displayed in Eq.(\ref{OMEGAPACA}) is reassuring given the
difference on the methods. It should be also mentioned that the so
called analytic scaling corrections have been neglected, which is, a
posteriori, seen to be a reasonable procedure~\cite{CORRECTIONS}.

Finally, we compare our estimate for the critical temperature
($T_{\mathrm c}=1.138(10)$) with that of Ref.~\cite{PACA}
($T_{\mathrm c}=1.156(15)$), the agreement being very good.

\begin{table}[b]
\vbox{
\begin{tabular}{lcdd}%\hline 
Authors             &   $J$-Distribution/Material & $\nu$   & $\eta$\\ 
\hline\hline
This Work ($(10,20)$ pair) & $\pm J$
&  2.15(15)  & --0.337(15)\\ \hline
Gunnarsson et al.~\cite{EXP}&${\mathrm Fe}_{0.5}{\mathrm Mn}_{0.5}{\mathrm TiO}_3$
&  1.69(15)  & --0.36(7)\\\hline
Palassini-Caracciolo~\cite{PACA}  & $\pm J$ 
&  1.8(2)    & -0.26(4)\\\hline
Marinari-Parisi-Ruiz-Lorenzo~\cite{MAPARU} & Gaussian 
&  2.00(15)  & --0.36(6)\\\hline
Berg-Janke~\cite{BERGJANKE} & $\pm J$ 
&  ---       & --0.37(4)\\\hline
Kawashima-Young~\cite{KAYOUNG} & $\pm J$ 
&  1.7(3)    & --0.35(5)\\\hline
I\~niguez-Parisi-Ruiz-Lorenzo~\cite{INPARU} & Gaussian 
&  1.5(3)    & ---     \\\hline
Ogielski~\cite{OGIELSKI} & $\pm J$ 
&  1.3(1)    & --0.22(5)\\\hline
Bhatt-Young~\cite{BY1} & $\pm J$ 
&  1.3(3)    & --0.3(2) \\\hline
Bhatt-Young~\cite{BY2} & Gaussian 
&  1.6(4)    & --0.4(2)%\\\hline
\end{tabular}
}
\caption[0]{Critical exponents from experiments and numerical
simulations.
\protect\label{table:sim}}
\end{table}

\section{Discussion and Conclusions}

We have obtained precise measures of the $\eta$ and $\nu$ exponents.
Moreover, we have done a study of the corrections to scaling, in
particular we have computed the value of the corrections-to-scaling
exponent $\omega$ in a good agreement with the value reported in
Ref.~\cite{PACA}. We remark that our statistical error for the $\eta$
exponent is 5 times smaller than the experimental error for this
exponent and 3 times less than the smallest statistical error found in
the literature. The fact that our estimates for the $(5,10)$ lattices
and the $(10,20)$ pair coincide within statistical errors gives us
some confidence on the smallness of the finite size effects, although
larger lattices will be needed to make sure that the systematic errors
are as small as the statistical ones.

Our comparison with the most recent experimental data~\cite{EXP} is
good. The difference between the $\nu$ exponent measured in experiment
and our reported value is $0.046(21)$, roughly two standard
deviations. We note that the $\nu$ exponent from numerical simulations
is systematically above experimental data. The difference for
the $\eta$ exponent is $0.023(71)$.

The clear crossing of the $\xi/L$ curves, for different lattice sizes,
supports heavily a finite temperature second order phase transition
and excludes a Kosterlitz-Thouless like scenario (a phase transition of
infinite order or equivalently a line of critical points below the
critical temperature).

Our results for the critical exponents for binary $J$'s agree well
with that obtained simulating Gaussian couplings~\cite{MAPARU}.
However the statistical error are still very large and we lack control
on the scaling corrections (completely in the Gaussian ``side''). 
It might be useful to study of the Gaussian EA model with the methodology
of this paper but unfortunately the SUE machine is not designed to
simulate Gaussian coupling.

\section{Acknowledgments}

We acknowledge discussions with E. Marinari and G. Parisi.
We thank partial financial support from CICyT (AEN97-1680, AEN97-1693,
AEN99-0990 and PB98-0842) and DGA (P46/97). V.M-M is a M.E.C. fellow.
The computations have been carried out using the RTNN machines
(Universidad de Zaragoza and Universidad Complutense de Madrid) and
the dedicated machine SUE (Universidad de Zaragoza).


\begin{thebibliography}{99}
\bibitem{SDE} One of the Schwinger-Dyson equations for a spin glass reads as:
$$
\frac{1}{V} \overline{\langle  \sum_i e^{-2 \beta \sigma_i S_i}
\rangle}=1 \; , \;\;\;  
S_i\equiv \sum_{j \,{\mathrm n.n.} \,i} J_{ij} \sigma_j \, ,
$$
where the sum ($\sum_{j \,{\mathrm n.n.} \,i}$) is performed over all
the six nearest neighbors of the site $i$.
\bibitem{HELICOIDAL} 
If we denote the coordinates of a given spin $(x,y,z)$ by 
$i\equiv x + y L + z L^2$, the
coordinates of the three nearest neighbors (in the three positive
directions) are given by: $i_1=(i+1) \,{\mathrm mod}\, L^3 $, $i_2=
(i+L) \,{\mathrm mod}\, L^3$ and $i_3= (i+L^2)\, {\mathrm mod}\, L^3$.
\bibitem{SIM} We have simulated the two dimensional XY model using the
Wolff single cluster algorithm~\cite{WOLFF}. We have done different
runs, near the critical point $\beta\simeq 1.11$, on lattice sizes
$L=16$, $32$ and $64$, using the spectral density method in the
analysis of the data.
\bibitem{CORRECTIONS} The leading correction to scaling is given by
the universal exponent $\omega$. However, the nonlinear dependence of
the scaling fields over the couplings of the Hamiltonian and the
analytic part of the free energy provided two sources of additional
scaling corrections.  Considering for instance the susceptibility, one
has that the leading exponents for the analytic scaling-corrections
(from each one of the above meantioned sources) are respectively:
$\gamma/\nu+d-1/\nu$ and $\gamma/\nu$. In our model they are
approximately $4.7$ and $2.3$ (resp.). These figures are much bigger
than our estimate for the $\omega$ exponent, which is 0.84(43), so
that we check, a posteriori, that they could be safely neglected. We
address to Ref.~\cite{SALASSOKAL} for a detailed study on the analytic
scaling-corrections.
\end{thebibliography}
\end{document}